\documentclass[11pt,twoside]{article}
\usepackage[margin=2cm]{geometry}

\usepackage{setspace}
\usepackage{float}
\usepackage{tikz}
\usepackage{hyperref}
\usepackage{nicematrix}
\usetikzlibrary{shapes.geometric, arrows}
\usepackage{algorithm2e}
\usepackage{amssymb }
\usepackage{graphicx}
\usepackage{breqn}
\usepackage{bbm}
\usepackage{synttree} 
\usepackage[utf8]{inputenc}
\usepackage{euscript}
\usepackage{amsmath,bm}
\usepackage{makecell}

\newcommand{\X}{\bm{X}}
\newcommand{\Y}{\bm{Y}}
\newcommand{\F}{\mathcal{F}}
\newcommand{\C}{\mathcal{C}}

\newcommand{\Xset}{\bm{\mathcal{X}}}

\DeclareRobustCommand{\frac}[3][0pt]{%
  {\begingroup\hspace{#1}#2\hspace{#1}\endgroup\over\hspace{#1}#3\hspace{#1}}}

\usetikzlibrary{matrix}
\usepackage{indentfirst}
\author{Zahra Gholamalian, Zeinab Maleki, MasoudReza Hashemi, Pouria Ramazi} 

\title{Detecting individual-level infections using sparse
group-testing through graph-coupled hidden Markov models} 
\makeatletter
\def\bbordermatrix#1{\begingroup \m@th
  \@tempdima 4.75\p@
  \setbox\z@\vbox{%
    \def\cr{\crcr\noalign{\kern2\p@\global\let\cr\endline}}%
    \ialign{$##$\hfil\kern2\p@\kern\@tempdima&\thinspace\hfil$##$\hfil
      &&\quad\hfil$##$\hfil\crcr
      \omit\strut\hfil\crcr\noalign{\kern-\baselineskip}%
      #1\crcr\omit\strut\cr}}%
  \setbox\tw@\vbox{\unvcopy\z@\global\setbox\@ne\lastbox}%
  \setbox\tw@\hbox{\unhbox\@ne\unskip\global\setbox\@ne\lastbox}%
  \setbox\tw@\hbox{$\kern\wd\@ne\kern-\@tempdima\left[\kern-\wd\@ne
    \global\setbox\@ne\vbox{\box\@ne\kern2\p@}%
    \vcenter{\kern-\ht\@ne\unvbox\z@\kern-\baselineskip}\,\right]$}%
  \null\;\vbox{\kern\ht\@ne\box\tw@}\endgroup}
\makeatother

\usepackage{biblatex}
\begin{document}
\setstretch{2}
\maketitle
\begin{abstract}
Identifying the infection status of each individual during the course of infectious diseases informs public health management. 
However, performing frequent individual-level tests may not be feasible.  
Instead, sparse and sometimes group-level tests are performed. 
Determining infection status of individuals using sparse group-level tests remains an open problem.
We tackled this problem by extending graph-coupled hidden Markov models with
individuals' infection statuses as the hidden states and the group test results as the observations.
We fitted the model to simulation datasets using the Gibbs sampling method. 
The model performed about 0.55 AUC for low testing frequencies and increased to 0.80 AUC in the case where the groups were tested every day. 
The model was separately tested in the daily basis case to predict the statuses over time and after 15 days of the beginning of the spread, which resulted in 0.98 AUC at day 16 and remained above 0.80 AUC until day 128. 
Therefore, although dealing with sparse tests remains unsolved, the results open the possibility of using initial group screenings during pandemics to accurately estimate individuals' infection statuses.

\end{abstract}

\section{Introduction} \label{sec_Introduction}


Mitigating pandemic diseases has been a challenge to governments.
Examples include Severe Acute Respiratory Syndrome (SARS), the multiple forms of influenza \cite{cook2008pandemic}, and COVID-19 \cite{ataguba2020covid}.
At the beginning of COVID-19 pandemic, its fast spread and many asymptomatic cases urged the public health management to detect infected individuals quickly and ask them to keep their social distancing or stay in quarantine \cite{ke2020fast}. 
Various datasets were collected on the potentially influential features on the disease spread \cite{gallo2021predictors, haratian2021dataset, hamzah2020coronatracker} and several machine-learning \cite{santosh2020covid, ramazi2021accurate} and hybrid data-driven and mechanistic models \cite{wang2022hypothesis, Ramazi3} were developed. 
Nevertheless, at the best, these models provided an estimation of the future \emph{total} number of infected cases, not individual cases.

Infection awareness at the individual level can be a self-controlling factor, decreasing the disease spread \cite{funk2009spread}. 
Yet infections may be asymptomatic.
Laboratory diagnostic methods allow the timely detection of asymptomatic carriers. 
However, at the beginning of a pandemic, frequent individual-level tests are infeasible due to the shortage of diagnostic tests and the limited capacity of laboratories, which was the case with COVID-19 \cite{udugama2020diagnosing}. 
A potential solution is to perform group-level testing, that is, to group and sample the individuals, mix the samples of the individuals in each group, and then perform testing on the resulting mixtures \cite{verdun2021group}. 
Back in World War II, statistician Robert Dorfman proposed a group-testing method to find army recruits infected with Syphilis \cite{dorfman1943detection}.

Group-testing strategies are time-efficient and save testing equipment.
However, to track the spread of a disease, group testings should be conducted over time. 
This results in a sequence of temporal group test results to be used to estimate individual infection statuses. 
To the best of our knowledge, this problem has not been studied before. 
The problem becomes more challenging in the desired more-realistic case where groups are tested sparsely over time rather than for example daily. 

The literature includes studies on estimating individual-level infection statuses based on individual-level disease observations, such as disease-specific symptoms or clinical testing results.
The authors in \cite{dong2012graph} studied the students living in a dormitory at the Massachusetts Institute of Technology (MIT), who reported their daily symptoms of flu, and whose contacts with each other were tracked via their mobile phone Bluetooth signals. 
They used the graph extension of coupled hidden Markov models to model the contacts between individuals and track their infection statuses. 
Taking daily personal symptom reports as the observation vector of the model, they performed inference and parameter learning by using the \textit{Gibbs\  sampling} algorithm. 
This approach needs daily personal symptom reports and individuals' high participation in the data collection.

In addition to the mobile phone Bluetooth signal collected data and the daily flu symptom reports, the work in \cite{fan2015hierarchical}, used some covariates referring to personal health and hygiene features such as weight, height, salads per week, sports per week, and smoking indicators to learn person-specific parameters of infection with the virus, along with tracking the spread of the disease.
A multivariate Gaussian distribution was used to capture the correlation between the personal covariates and the amount of personal susceptibility to the infection. 
By applying the Expectation Maximization (EM) algorithm, every individual's health statuses at every timestamp of the data was inferred.
This approach has a high accuracy at the expense of being highly data demanding.

All previous work used individual-level observations available through time, to estimate individual-level infection statuses.
To tackle the problem of estimating individual-level infection statuses from only group-level observations and the individuals' contacts over time, we extended the graph hidden Markov model in \cite{dong2012graph} where individuals are grouped into ``families'' and a single observation is available for the whole family rather than each individual in that family.
The observation may also be available only at some time steps. 
We used Gibbs sampling to simultaneously learn the model parameters and infer the model estimation of individual-level health statuses over synthesis datasets.
The stochastic Gibbs sampling approach may be particularly an appropriate alternative to deterministic algorithms such as Expectation-Maximization (EM) when the network size is large \cite{casella1992explaining}. 
This method starts by generating a sample for the latent variables from some initial distribution, then iterates over each of the latent variables and samples a new value for each one, conditioned on the current sample of all the others \cite{koller2009probabilistic}. 

The rest of the paper is organized as follows. 
In section \ref{sec_ProblemFormulation}, we describe the problem formulation in detail. In Section \ref{sec_Methods}, we explain our approach to parameter learning and inference of the model, followed by some evaluation strategies and data simulation. 
We report the results of conducting several experiments on our simulation data in Section \ref{sec_Results} and provide conclusions and future directions in Section \ref{sec_Conclusion}.

\section{Problem formulation}   \label{sec_ProblemFormulation}
Consider a population of $I$ interacting individuals who are susceptible to a certain infectious disease and may become infected  over time $t = 0,1,\ldots, T$. 
Each individual $i$ has a health status $X_{i,t}$ at time $t$, which is either 0 (healthy) or 1 (infected). 
The individuals are grouped into several ``families'', which may represent for example, actual families, housemates, or roommates. 
The health statuses are unknown and are revealed by some diagnosis tests. 
Instead of testing the individuals separately and at every time step, group testings are performed for the families and sparsely at certain time steps. 
A family is considered as infected if at least one of the family members is infected.
The test result of family $f$ at time $t$ is indicated by $Y_{f,t}$, which is either 0 (negative) or 1 (positive). 

Denote the probability of a positive test result for a healthy and infected individual by $\theta_0$ and $\theta_1$ respectively, where we assume $\theta_0<\theta_1$.
Then the probability that family $f$ with size $n_f$ is tested positive, given  that $n'_{f,t}$ of its members are infected  at time $t$ takes the following distribution:
\begin{equation}    \label{eq_emissionProbability}
    P\Big(Y_{f,t}=1 \mid  \sum_{i\in f} X_{i,t} = n'_{f,t}\Big) \sim \text{Bernoulli}\Big({((n_f-n'_{f,t})\theta_0+n'_{f,t}\theta_1)}/{n_f}\Big).
\end{equation}
The individuals become infected via their contacts in and outside the population. At every time $t$, in addition to all of her family members, each individual $i$ interacts with her non-family contacts denoted by $\mathcal{C}_{i,t}\subseteq\{1,\ldots,I\} $ and possibly some individuals outside the considered population.
A susceptible individual becomes infected with probability $\beta_f$ by contacting with a family member, with probability $\beta$ by a non-family member  within the considered population, and  with probability $\alpha$ by a person outside the considered population. 
If the susceptible individual $i$ at time $t$ has $n'_{f,t}$ infected family contacts and $c'_{i,t}$ infected non-family contacts, she becomes infected with the overall probability 
$$
    1-(1-\alpha)(1-\beta)^{c'_{i,t}}(1-\beta_f)^{n'_{f,t}}
$$
and remains susceptible otherwise. 
The infection probabilities $\alpha$, $\beta$, and $\beta_f$ are assumed to be small, allowing for the above term to be approximated as $\alpha+\beta{c'_{i,t}}+\beta_f{n'_{f,t}}$. 
An individual that is infected at time $t$ recovers at time $t+1$ with probability $\gamma$ and remains infected with probability $1-\gamma$. 
Hence the \textit{transition probability} from state $X_{i,t}$ to $X_{i,t+1}$ is captured by the following matrix:
\begin{equation}    \label{eq_transitionMatrix}
    \begin{bbordermatrix}
    {  &  \scalebox{0.7}{%
    $X_{i,t+1}=0$} & \scalebox{0.7}{%
    $X_{i,t+1}=1$} \cr
    \scalebox{0.7}{%
      $X_{i,t}=0$} & 1-\alpha-\beta{c'_{i,t}}-\beta_f{n'_{f,t}}   & \alpha+\beta{c'_{i,t}}+\beta_f{n'_{f,t}} \cr
      \scalebox{0.7}{%
      $X_{i,t}=1$} &\gamma& 1-\gamma  }.\qquad
    \end{bbordermatrix}
\end{equation}
All parameters $\alpha, \beta, \beta_f, \gamma, \theta_0, \theta_1$ are unknown.
We assume the infection probability from an infected family member is higher than an infected non-family member within the network, which in turn is higher than from an infected individual out of the network, i.e., 
$\alpha<\beta<\beta_f$.
Stack all individuals' health statuses at time $t$ to obtain the state vector $\X_{t}$ = $(X_{1,t},...,X_{I,t})^\top$.
The goal is to find the final health state $\X_T$ using the family test results. 

\section{Methods}   \label{sec_Methods}
\subsection{Model}
As each individual's health status depends on that of other individuals only at the previous time step, it follows that $\X_{t}$ is a first order Markov chain \cite{murphy2012machine}; that is,
\begin{equation*} \label{MarkovAssumption}
    \X_{t} \perp\X_{t'} \mid \X_{t-1}\quad \forall t'<t-1.
\end{equation*}
Indeed, those ``other individuals'' are limited to the contacts of the individual at the previous time step, resulting in 
\begin{equation} \label{healConditionIndependenceAtTimeT1}
    X_{i,t} \perp \X_{t'} \mid 
    \Xset_{i,t-1}
    \quad \forall t'< t-1,
\end{equation}
where $\Xset_{i,t}$ is the set of health statuses of all individual $i$'s contacts at time $t$, including family and non-family, i.e.,
$
    \Xset_{i,t}
    =
    \{X_{j,t}\mid j\in \F_i \cup\mathcal{C}_{i,t} \},
$
where $\F_i$ is the set of family members of individual $i$.
Clearly, the independence extends to the health statuses of the non-contacted individuals as well. 
That is, knowing the statuses of individual $i$'s contacts at time $t-1$, the statuses of the other individuals at that time does not provide further information about individual $i$'s current health status: 
\begin{equation} \label{healConditionIndependenceAtTimeT2}
     X_{i,t} \perp 
    \X_{t-1}\setminus\Xset_{i,t-1} \mid 
    \Xset_{i,t-1}.
\end{equation}
The special case of $t=0$ implies that the health statuses are initially mutually independent, i.e., 
$
    X_{i,0} \perp X_{j,0}
$    
for all $i\neq j$.

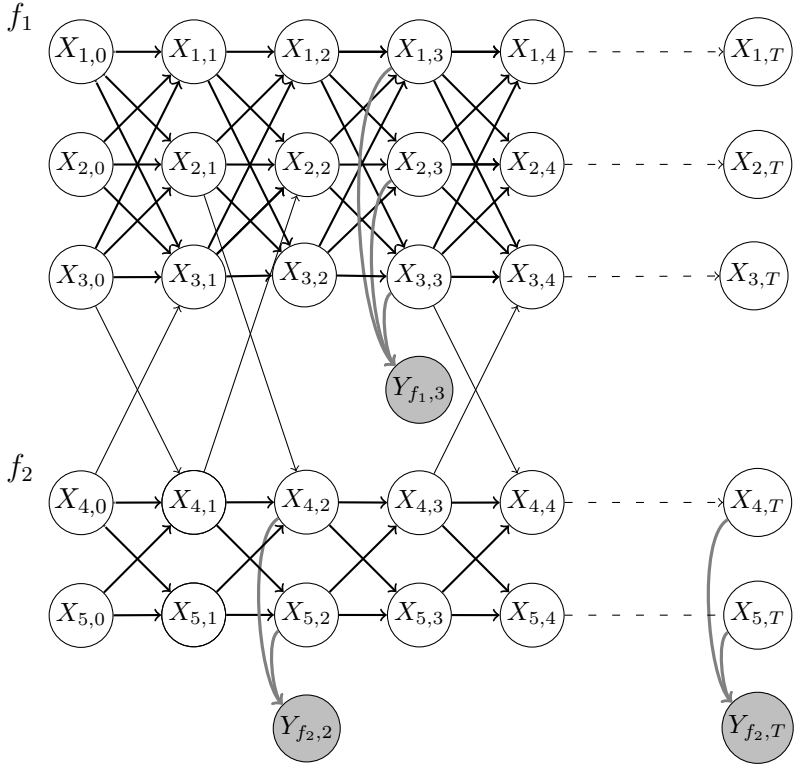
\begin{figure}[!h]   \label{figure_graphHMMExample}
    \begin{tikzpicture}\tikzstyle{every node}=[draw,shape=circle, minimum size = 0.2cm, inner sep=1pt];
    \node (v0) at (   0:0) {\color{white}\small$X_{1,0}$};
    \node (v1) at (  -90:1.5) {\small$X_{2,0}$};
    \node (v2) at (   0:1.5) {\small$X_{1,1}$};
    \node (v3) at (  -45:2.12) {\small$X_{2,1}$};
    \node (v4) at (   0:3) {\small$X_{1,2}$};
    \node (v5) at (  -26.56:3.36) {\small$X_{2,2}$};
    \node (v6) at (   0:6) {\small$X_{1,4}$};
    \node (v7) at (  -14.03:6.18) {\small$X_{2,4}$};
    \node (v8) at (   0:4.5) {\small$X_{1,3}$};
    \node (v9) at (  -18.43:4.74) {\small$X_{2,3}$};
    \node (v10) at (  -33.69:5.4) {\small$X_{3,3}$};
    \node (v11) at (   0:9) {\small$X_{1,T}$};
    \node (v120) at (  -9.46:9.12) {\small$X_{2,T}$};
    \node (v130) at (  -18.43:9.43) {\small$X_{3,T}$};
    \node (v02) at (  -90:3) {\small$X_{3,0}$};
    
    \node (v22) at (   -63.43:3.35) {\small$X_{3,1}$};
    \node (v42) at (   -45:4.2) {\small$X_{3,2}$};
    \node (v62) at (   -26.56:6.7) {\small$X_{3,4}$};
    \node (v72) at (  -45:6.36)[fill=gray!50] {\small$Y_{f_1,3}$};
    
    \node (v03) at (  -90:6) {\color{white}\small$X_{4,0}$};
    \node (v13) at (  -90:7.5) {\small$X_{5,0}$};
    \node (v23) at (   -75.96:6.18) {\color{white}\small$X_{4,1}$};
    \node (v33) at (  -78.7:7.64) {\color{white}\small$X_{5,1}$};
    \node (v43) at (    -63.43:6.7) {\small$X_{4,2}$};
    \node (v53) at ( -68.2:8.07) {\small$X_{5,2}$};
    \node (v63) at (   -45:8.48) {\small$X_{4,4}$};
    \node (v73) at (  -51.34:9.6) {\small$X_{5,4}$};
    \node (v030) at (  -53.13:7.5) {\small$X_{4,3}$};
    \node (v1030) at (  -59.03:8.74) {\small$X_{5,3}$};
    \node (v230) at (   -75.96:6.18) {\small$X_{4,1}$};
    \node (v330) at (  -78.7:7.64) {\small$X_{5,1}$};
    \node (v2030) at (   -33.69:10.81) {\small$X_{4,T}$};
    \node (v3030) at (  -39.8:11.71) {\small$X_{5,T}$};
    
    
    \node (v24) at (  -71.57:9.49)[fill=gray!50] {\small$Y_{f_2,2}$};
    \node (v304) at (  -45:12.72)[fill=gray!50] {\small$Y_{f_2,T}$};
    
    \draw[thick,->] (v0) -- (v2);
    \draw[thick,->] (v2) -- (v4);
    \draw[thick,->] (v0) -- (v22);
    \draw[thick,->] (v02) -- (v2);
    \draw[thick,->] (v2) -- (v42);
    \draw[thick,->] (v22) -- (v4);
    \draw[thick,->] (v4) -- (v8);
    \draw[thick,->] (v8) -- (v6);
    
    \draw[thick,->] (v1) -- (v3);
    \draw[thick,->] (v3) -- (v5);
    \draw[thick,->] (v5) -- (v9);
    \draw[thick,->] (v9) -- (v7);
    
    \draw[thick,->] (v10) -- (v7);
    \draw[thick,->] (v9) -- (v62);
    \draw[thick,->] (v9) -- (v6);
    \draw[thick,->] (v8) -- (v7);
    
    \draw[thick,->] (v8) -- (v6);
    \draw[thick,->] (v9) -- (v7);
    \draw[thick,->] (v10) -- (v62);
    \draw[thick,->] (v8) -- (v62);
    \draw[thick,->] (v10) -- (v6);
    
    \draw[thick,->] (v0) -- (v3);
    \draw[thick,->] (v1) -- (v2);
    \draw[thick,->] (v2) -- (v5);
    \draw[thick,->] (v3) -- (v4);
    
    \draw[thick,->] (v4) -- (v8);
    \draw[thick,->] (v4) -- (v10);
    \draw[thick,->] (v5) -- (v8);
    \draw[thick,->] (v42) -- (v9);
    
    \draw[thick,->] (v42) -- (v8);
    \draw[thick,->] (v4) -- (v9);
    \draw[thick,->] (v5) -- (v10);
    \draw[thick,->] (v22) -- (v5);
    
    \draw[thick,->] (v02) -- (v3);
    \draw[thick,->] (v1) -- (v22);
    \draw[thick,->] (v3) -- (v42);
    \draw[thick,->] (v22) -- (v5);
    \draw[loosely dashed,->] (v6) -- (v11);
    \draw[loosely dashed,->] (v7) -- (v120);
    \draw[loosely dashed,->] (v62) -- (v130);

    \draw[very  thick,gray,->] (v8) to [out=210, in=135, looseness=.5](v72);
    \draw[very  thick,gray,->] (v9) to [out=210, in=135, looseness=.5](v72);
    \draw[very  thick,gray,->] (v10) to [out=210, in=135, looseness=.5](v72);
    
    \draw[very  thick,gray,->] (v2030) to [out=210, in=135, looseness=0.5](v304);
    \draw[very  thick,gray,->] (v3030) to [out=210, in=135, looseness=0.5](v304);
    
    \draw[thick,->] (v02) -- (v22);
    \draw[thick,->] (v22) -- (v42);
    \draw[thick,->] (v42) -- (v10);
    \draw[thick,->] (v10) -- (v62);
    
    \draw[thick,->] (v53) -- (v030);
    \draw[thick,->] (v43) -- (v1030);
    \draw[thick,->] (v030) -- (v73);
    \draw[thick,->] (v1030) -- (v63);
    
    \draw[thick,->] (v03) -- (v23);
    \draw[thick,->] (v23) -- (v43);
    \draw[thick,->] (v43) -- (v030);
    \draw[thick,->] (v53) -- (v1030);
    \draw[thick,->] (v1030) -- (v73);
    \draw[thick,->] (v030) -- (v63);
    \draw[loosely dashed,->] (v63) -- (v2030);
    \draw[thick,->] (v13) -- (v33);
    \draw[thick,->] (v33) -- (v53);
    \draw[loosely dashed] (v73) -- (v3030);

    \draw[very thick, gray,->] (v43) to [out=210, in=135, looseness=.5](v24);
    \draw[very thick, gray,->] (v53) to [out=210, in=135, looseness=.5](v24);

    \draw[thick,->](v03)--(v33);
    \draw[thick,->](v13)--(v23);
    \draw[thick,->](v23)--(v53);
    \draw[thick,->](v33)--(v43);
    

    \node[label={[left=14]:\large $f_{1}$},draw=white] () at (0,0) {\color{black}$X_{1,0}$};
    \node[label={[left=14]:\large $f_{2}$},draw=white] () at (0,-6) {\color{black}$X_{4,0}$};
    
    \draw[thin,->](v02)--(v23);
    \draw[thin,->](v23)--(v5);
    
    \draw[thin,->](v03)--(v22);
    \draw[thin,->](v3)--(v43);
    
    \draw[thin,->](v10)--(v63);
    \draw[thin,->](v030)--(v62);
    
    \end{tikzpicture}
    \caption{\textbf{Representation of the disease spread.} The population consists of 5 individuals, where the first 3 are from family $f_1$, and the last two are from family $f_2$. 
    The members from each family have common observable test nodes. 
    Contacts between family members persist over time, but non-family contacts may change at each time step. 
    Family $f_1$ was tested at time step 3, family $f_2$ was tested at time steps 2 and $T$.}
\end{figure}

Assumptions \eqref{healConditionIndependenceAtTimeT1} and \eqref{healConditionIndependenceAtTimeT2} provide a local Markov independent assumption for every variable $X_{i,t}$ \cite{koller2009probabilistic}.
Hence, the infection spreading dynamics can be modeled by using an extended \emph{hidden Markov Model (HMM)}, called \emph{graph-coupled HMM (GCHMM)} \cite{dong2012graph} where rather than a linear connection, the hidden states $X_{i,t}$ are connected according to the individuals' contacts over time.
More specifically, if individuals $i$ and $j$ met at time $t$, then there is a link from $X_{j,t}$ to $X_{i,t+1}$ and from $X_{i,t}$ to $X_{j,t+1}$ (Figure~\ref{figure_graphHMMExample}).
Family members are assumed to meet at every time step. 
Moreover, corresponding to each family $f$ tested at time $t$, every $X_{i,t}$ for family member $i$ is linked to the test node $Y_{f,t}$. 
This makes the family members' health statuses dependent on the test result, and once their health statuses are known, the test result is independent of all other nodes in the graph.

This extended GCHMM is only partially observed, as the group testing results are not available for every family at every time. 
Existing methods for learning the parameters of the HMM and consequently estimating the hidden states are limited to the fully observable case where every individual or family is tested at every time step.  
The sparsity in the observation data is the main challenge in this problem. 

\subsection{Parameter learning and inference}
To estimate $\X_T$, we follow an algorithm similar to the one in \cite{dong2012graph}; that is, after initializing the parameters and initial states, we iteratively estimate all hidden states and update the parameters until they all converge.
More specifically, we perform the following steps:

\textbf{Step 0. Initialization.} 
We assume all individuals are initially susceptible, i.e., $\X_0 = \bm{0}$, where $\bm{0}$ is the vector of all zeros. 
For the model parameters, we consider a Beta distribution for their priors:
\begin{equation}    \label{BetaHyperParameters}
    \begin{aligned}
         &\alpha\sim\text{Beta}(a_{\alpha},b_{\alpha}),\  
        \beta\sim\text{Beta}(a_{\beta},b_{\beta}),\ 
        \beta_f\sim\text{Beta}(a_{\beta_f},b_{\beta_f}),\\ 
        &\gamma\sim\text{Beta}(a_{\gamma},b_{\gamma}),\ 
        \theta_0\sim\text{Beta}(a_{\theta_0},b_{\theta_0}),\ 
        \theta_1\sim\text{Beta}(a_{\theta_1},b_{\theta_1}),
    \end{aligned}
\end{equation}
where, $a_\alpha, b_\alpha, a_\beta, b_\beta, a_{\beta_f}, b_{\beta_f}, a_\gamma, b_\gamma, a_{\theta_0}, b_{\theta_0}, a_{\theta_1}, b_{\theta_1}$ are hyperparameters and are randomly initialized. 
We accordingly sample the parameters.

\textbf{Step 1. Estimating $\X$.}
Let $\X$ be the set of all hidden states and $\Y$ be the set of all family tests--both over the whole time horizon $0,1,\ldots,T$.
Recall that we already have the values of $\Y$.
At this step, we estimate $\X$ based on $\Y$.
Due to the possibly large size of $\X$, rather than an exact inference, we perform an approximate inference by sampling. 

First, we generate one sample of $\X$ using the \emph{forward sampling}; that is, using $\X_0$ from the previous step, we sample $\X_1$ using the transition matrix \eqref{eq_transitionMatrix}, where the infection parameters are initialized in the previous step, and similarly we sample $\X_2$ based on $\X_1$ and so on. until we have sampled the entire $\X$.

Next, we ``improve'' the samples via \emph{Gibbs sampling}, that is, in turn, for $t=1,\ldots,T$, and for all individuals $i$, to re-sample $X_{i,t}$ based on all variables but $X_{i,t}$, i.e., $\{\X,\Y\}\setminus \{X_{i,t}\}$, and then update the value of $X_{i,t}$ in $\X$. 
The re-sampling is done based on the following conditional probability:
\begin{equation}   \label{eq_GibbsSampling1}
    \begin{aligned}
         P(X_{i,t}\mid\{\X,\Y\}\setminus \{X_{i,t}\})
         & = \frac{P(\X,\Y)}{\sum_{X_{i,t}=0,1}P(\X,\Y)}.
    \end{aligned}
\end{equation}
As the joint probability of all of the variables factorizes according to the GCHMM, that is, the product of the probability of each node conditioned on its parents, we obtain 
\begin{equation*}
    P(\bm{X}, \bm{Y})
    =\prod_{i=1}^{I}\prod_{t=1}^{T}
    P(X_{i,0})P(X_{i,t}\mid \Xset_{i,t-1})
    \prod_{f=1}^{F}\prod_{t_f\in\mathcal{T}_f}\!\!\!
    P(Y_{f,t_f}\mid \{X_{j,t_f}{,j} \in f\}),
\end{equation*}
where $\mathcal{T}_f\subseteq \{0,1,\ldots,T\}$ is the set of the time steps that each family $f$ is tested, and 
the term $P(X_{i,t}\mid \Xset_{i,t-1})$ is computed based on \eqref{eq_transitionMatrix} and 
$$ 
    P(Y_{f,t_f}\mid \{X_{i,t_f}{,i\in f}\})
    = P\Big(Y_{f,t}=1 \mid  \sum_{i\in f} X_{i,t} = n'_{f,t}\Big),
$$ 
which is computed based on \eqref{eq_emissionProbability}. 
Hence, \eqref{eq_GibbsSampling1} simplifies to only the terms including $X_{i,t}$:
\begin{equation}   \label{eq_GibbsSampling2}
    \begin{aligned}
         P&(X_{i,t}\mid\{\X,\Y\}\setminus \{X_{i,t}\})\\
         & = \frac{\scalebox{0.9}{$
                P(X_{i,t}\mid \Xset_{i,t-1})
                \left[\prod_{j\in\C_{j,t}}
                P(X_{j,t+1}\mid \Xset_{j,t})\right]
                P(Y_{f,t_f}\mid \{X_{j,t_f}\mid j\in f\})$}}%
            {\scalebox{0.9}{$
                \sum_{X_{i,t}=0,1}
                P(X_{i,t}\mid \Xset_{i,t-1})
                \left[\prod_{j\in\C_{j,t}}
                P(X_{j,t+1}\mid \Xset_{j,t})\right]
                P(Y_{f,t_f}\mid \{X_{j,t_f}\mid j\in f\})$}},
    \end{aligned}
\end{equation}
where $f$ is the family of individual $i$. 
See the appendix for an example of how this equation is calculated. 

Repeating the same process for all of the entries of $\X$, one at a time, we generate a new sampled value for each entry given the current values of others.
We then repeat the whole process, i.e., to re-sample every entry of $\X$, until the changes in $\X$ become smaller than a specified threshold, implying that the distribution from which we generate each sample has converged to its posterior.

\textbf{Step 2. Updating the parameters.}
Based on the estimated states $\X$, we update the model parameters in this step.
The parameters $(a_Z,b_Z)$ of the Beta distributions of each parameter $Z\in\{\alpha,\beta,\beta_f,\gamma, \theta_0,\theta_1\}$ in \eqref{BetaHyperParameters} are updated to the posterior parameters $(a'_Z,b'_Z)$ according to the linear relationship \cite{murphy2012machine}:
\begin{equation*}
    a'_Z=a_Z+n_Z',\qquad
    b'_Z=b_Z+n_Z-n_Z',
\end{equation*}
where $n'_Z$ is the number of $(i,t)$ instances where individual $i$ \textbf{was} infected (resp. recovered or remained susceptible) according to the incidence corresponding to the parameter $Z$, and $n_Z$ is the total possible number of $(i,t)$ instances where individual $i$ \textbf{could have been} infected (resp. recovered or remained susceptible) according to the incidence corresponding to $Z$.
More specifically, 
\begin{itemize}
    \item $n'_\alpha$ is the number of instances where an individual is infected from outside the network, and $n_\alpha$ is the total number of healthy individuals, i.e., the number of $(i,t)$ instances where $X_{i,t}=0$;
    \item $n'_\beta$ is the number of instances where an individual is infected by a non-family member inside the network, and $n_\beta$ is the number of $(i,t)$ instances, where a non-family contact of individual $i$ at time $t-1$ was infected at time $t-1$;
    \item $n'_{\beta_f}$ is the number of instances where an individual is infected by a family member, and $n_{\beta_f}$ is the number of $(i,t)$ instances, where a family member of individual $i$ was infected at time $t-1$;
    \item $n'_{\gamma}$ is the number of instances where an infected individual recovers, and $n_{\gamma}$ is the number of instances $(i,t)$ where individual $i$ is infected at time $t$;
    \item $n'_{\theta_0}$ is the number of $(i,t)$ instances where individual $i$ is susceptible but her family test at time $t$ is positive, and $n_{\theta_0}$ is the number of $(i,t)$ instances where individual $i$ is susceptible at time $t$;
    \item $n'_{\theta_1}$ is the number of $(i,t)$ instances where individual $i$ is infected and her family test at time $t$ is positive, and $n_{\theta_1}$ is the number of $(i,t)$ instances where individual $i$ is infected at time $t$.
\end{itemize}
The numbers $n'_{\alpha}, n'_{\beta}, n'_{\beta_f}$ depend on the source of infection which is unknown even after sampling the whole health statuses $\X$.  
Hence, we define the auxiliary variable \emph{infection origin} $O_{i,t}$ that indicates the infection origin of individual $i$ at time  $t\geq 1$. 
Should the healthy individual $i$ at time $t$ remain healthy at time $t+1$, $O_{i,t}=0$. 
But if she becomes infected at time $t+1$, for the infection origin from outside of the considered population, $O_{i,t}=1$, inside the family  $O_{i,t}=2$, and outside the family but inside the population, $O_{i,t}=3$. 
Then $O_{i,t}$ follows the categorical conditional probability distribution specified by 
\begin{equation}
    \begin{aligned}
    & P(O_{i,t}=1\mid X_{i,t}=0,X_{i,t+1}=1)=\frac{\alpha}{\alpha+\beta c'_{i,t}+\beta_f n'_{i,f}},\\
    & P(O_{i,t}=2\mid X_{i,t}=0,X_{i,t+1}=1)=\frac{\beta}{\alpha+\beta c'_{i,t}+\beta_f n'_{i,f}},\\
    & P(O_{i,t}=3\mid X_{i,t}=0,X_{i,t+1}=1)=\frac{\beta_f}{\alpha+\beta c'_{i,t}+\beta_f n'_{i,f}}.
    \end{aligned}
\end{equation}
After sampling the value of $O_{i,t}$ for every individual $i$ at every time $t$, the numbers $n'_{\alpha}, n'_{\beta}, n'_{\beta_f}$ are readily obtained by counting the number of instances corresponding to each case. 

After updating the hyperparameters, the parameters are re-sampled from \eqref{BetaHyperParameters}. 
If the hyperparameters do not satisfy the conditions $\alpha<\beta<\beta_f$ and $\theta_0<\theta_1$, we re-sample the corresponding hyperparameters.

\textbf{Step 3. Repetition.}
We repeat Steps 1 \& 2 until the change in the parameters in Step 2 falls short of a specified threshold or a specified maximum number of iterations has reached. 

These three steps are summarized in Algorithm \ref{algorithm1}.

\RestyleAlgo{ruled}
\SetKwComment{Comment}{/* }{ */}
\begin{algorithm}[H]    \label{algorithm1}
    \caption{Our Gibbs sampling algorithm}\label{alg:one}
    \KwData{$\Y$ and $\F_i$ and $\C_{i,t}$ for every individual $i$ and time $t$
    }
    \KwResult{Health\ state\ matrix\  $X$,\ Optimized\ parameters}
    \textbf{Initialization:}\ randomly initialize the hyperparameters, then from \eqref{BetaHyperParameters} sample initial parameters 
    $\alpha,\beta,\beta_f,\gamma,\theta_0,\theta_1$, where $\alpha<\beta<\beta_f$ and $\theta_0<\theta_1$;
    obtain the initial value of $\bm X$ by setting 
    $\bm X_{0}$ to $\bm{0}$ and sampling $\bm X_1, \ldots,\bm X_n$ using forward sampling and transition matrix \eqref{eq_transitionMatrix}.\\
      \Repeat{the parameters and $X$ converge}{
        \Repeat{$\bm X$ converges with regard to the current parameters}{
         re-sample $\bm X$ using Gibbs sampling\;
         }
         {update parameters according to \textbf{Step 2}}\;
      }
\end{algorithm}
\subsection{Experiment setup}
To evaluate the model, we generated two synthesized datasets, one over 360, where the families were sparsely tested, and the other over 128 days, where every family was tested every day (Table \ref{table:1}).
\begin{table}[h!]
    \centering
    \caption{Simulation data characteristics.}
    \begin{tabular}{|c|c|c|c|c|} 
     \hline
     \thead{Dataset\\ } & \thead{Population\\ size} & 
     \thead{Number\\ of families} & \thead{Number\\ of days} & $\mu$\\ 
     \hline
        1 & 100 & 33 & 360 & 1, 2, 3, 4, 5, 6, 12, 25, 52, 360 \\
        2 & 64 & 15 & 128 & 128\\ 
     \hline
    \end{tabular}
    \label{table:1}
\end{table}
We assumed that the number of tests during a year for each family was fixed to $\mu$.
We constructed a random social network for each population that determines for each individual $i$, her contacts $\C_{i,t}$ over time $t$. 
We did this by generating a random adjacency matrix for each time step $t$. 
Next, we clustered the individuals into the pre-specified number of families by first specifying the size of each family, limited to 5, then randomly selecting the individuals to be in that family.
We randomly chose the parameters $\alpha, \beta, \beta_f$ satisfying $\alpha<\beta<\beta_f$ and limited to the interval $(0,0.005]$, $\theta_0$ and $\theta_1$ satisfying $\theta_0<\theta_1$ and limited to the intervals $[0.01,0.03]$ and $[0.8,1)$ respectively, and $\gamma$ from the interval $[0.1,0.5]$.
We then generated the health statuses $\X$ by setting $\X(0)=\bf 0$ and sampling $\X(t+1)$ from $\X(t)$ for $t=0,\ldots,T-1$ using the transition matrix.
Every family was tested $\mu$ days that were chosen randomly from the whole $T$ days.
We then sampled the observation values $Y$ according to \eqref{eq_emissionProbability}. 

For each value of $\mu$, we passed the observations $\Y$, members of the families, and contacts over time to our algorithm and estimated the values of $\X$, indicating the infection probabilities for each individual over time.
Since the infection rates were chosen to be small, the resulting number of infected individuals in $\X$ would be much less than the number of healthy ones, resulting in an unbalanced dataset. 
Hence, accuracy might not properly assess the model performance.
Instead, we used the Receiver Operating Characteristic (ROC) curve and Area Under the Curve (AUC) measure \cite{fawcett2006introduction}.
This experiment was performed using the first dataset.

We performed a second experiment on the second dataset, where the model was evaluated through different time intervals.
As all individuals were initially healthy, we considered an initial period of 15 days for the progress of the disease to reach the epidemic stage and estimated the health statuses at time steps $t=16, 24, 32, \ldots, 128$.
At each time step $t$, only the observation values $Y$ prior and up to time $t$ were used to estimate the health statuses $\X(0),\X(1)\ldots,\X(t)$, that is, to set $T$ to $t$ in the algorithm. 
Then the AUC of the estimated $\X(t)$ values was obtained by comparing them to the ``true'' values that were obtained initially from the simulations. 
This experiment mimics the real-world situation where the goal is to predict individuals' current statuses based on current and past test results--future results are unavailable and predicting the individuals' past statuses is not of interest.
The reader may refer to \cite{ramazi2021predicting} for more information on this ``temporal partitioning'' of the dataset in the training and testing datasets.

Similar to \cite{dong2012graph}, we compared our model with a support vector machine (SVM).
We used the number of contacts (including family and non-family) with infected individuals at times $t-1$, $t$, and $t+1$ as the features for every individual $i$ at every time $t$. 
Using the \texttt{SVM} package in Python, we trained the SVM with a linear kernel on 80\% of the above data and used the remaining 20\% percent for the test. 
The missing values of the observation variable $\Y$ in the first experiment were imputed prior to the training by using \texttt{SimpleImputer} with the ``mean'' strategy from the \texttt{scikit-learn} package in Python.
\section{Results}   \label{sec_Results}
\begin{figure}[h!] \label{mu_Result}
    \centering
   \scalebox{0.7}{%
    \graphicspath{ {./mu_result/} }
\includegraphics{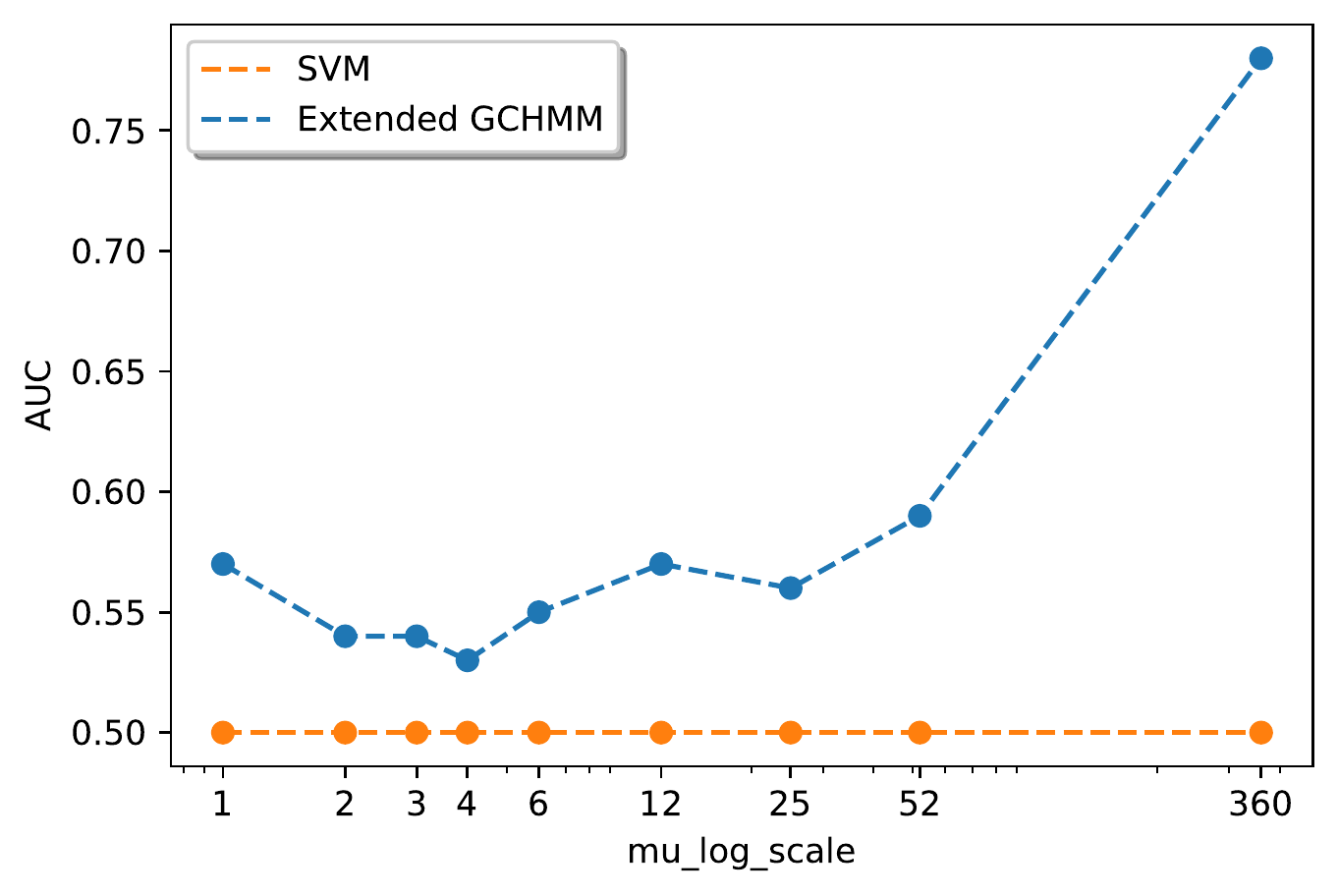}}
    \caption{\textbf{Model performance with respect to the number of family tests per family.} 
    The graph shows the AUC of the model predictions of all individuals' health statuses through 360 days against different values of $\mu$ for both the extended GCHMM and an SVM.}
\end{figure}

The model estimated the health statuses of the individuals over the 360 days in the first experiment with 0.57 AUC when each family was tested only once in this time interval (Figure~\ref{mu_Result}). 
The performance fluctuated as $\mu$ increased until it exceeded 0.6 AUC at $\mu=52$ and reached 0.8 AUC when each family was tested daily. 
The SVM model performed close to a random classifier, that is an AUC of about 0.5, regardless of the value of $\mu$. 

\begin{figure}[h!]\label{time_Result}
    \centering
   \scalebox{0.7}{%
    \graphicspath{ {./mu_result/} }
    \includegraphics{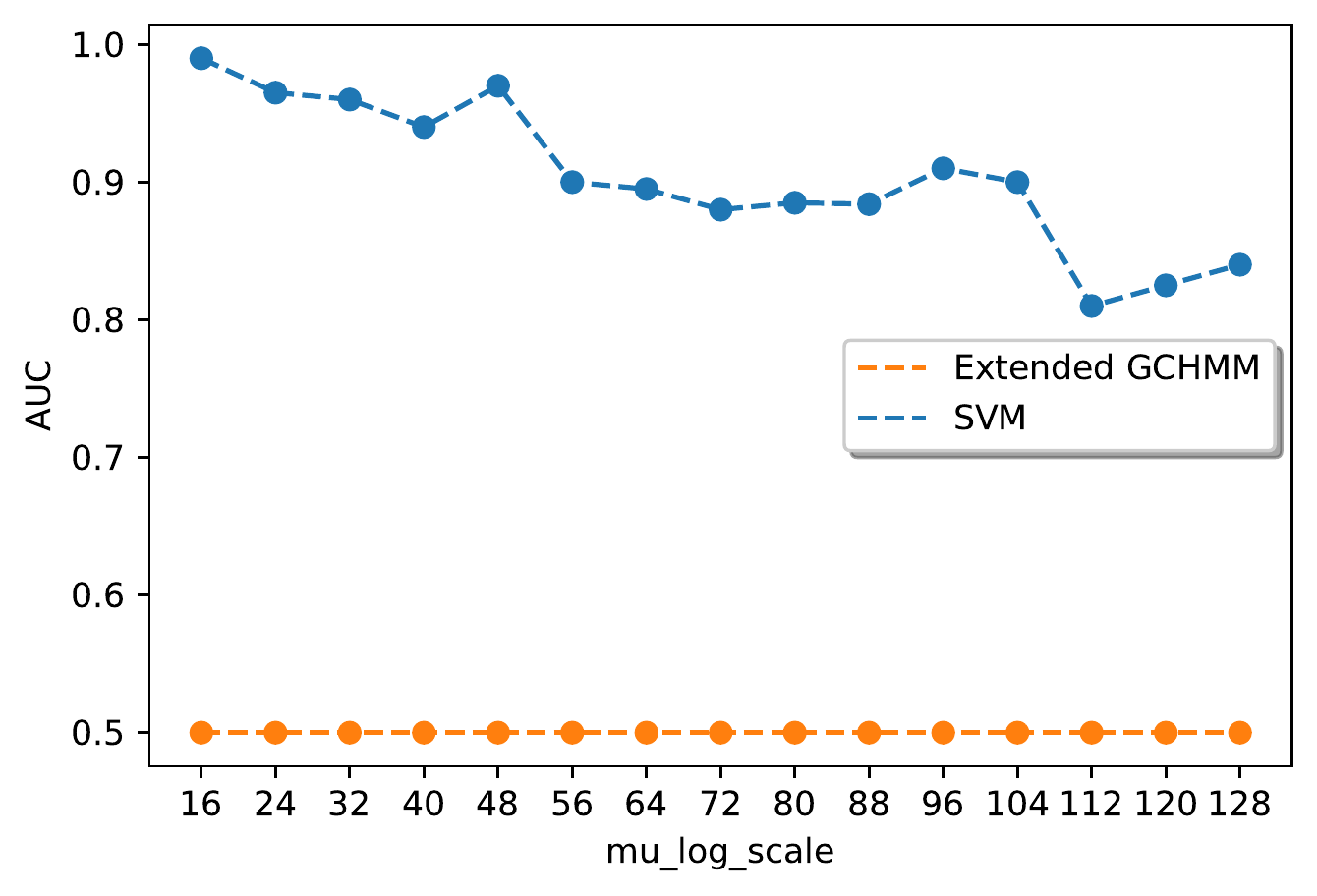}}
    \caption{\textbf{Model performance over time.}
    The graph shows the AUC performance of the extended GCHMM and an SVM in prediction of all individuals' health statuses through the 128 days in the second experiment. 
    Each family was tested every day in this experiment. 
    At each time step, the test results up to only that time were used for making predictions.
    }
\end{figure}

When all families were tested daily, the model predicted the individuals' statuses almost perfectly in the beginning days of the pandemic (after 15 days), and the performance deteriorated over time as a longer history of family tests are used to make predictions (Figure~\ref{time_Result}).
The SVM again performed 50\% AUC.
\section{Conclusion}    \label{sec_Conclusion}
During the spread of infectious diseases, especially at the early stages, diagnosis tests are either limited or yet not developed.
Sparse group tests are a potential solution to reduce the number of required tests. 
However, they do not reveal individual-level infection statuses needed for self-quarantine, contact tracing, and other follow-up steps in case of a positive test result. 
We tackled this problem by extending the GCHMM where individuals are divided into several groups, and rather than each individual, each group has an observation (emission) node, which is also only partially observed at some time steps. 
Using the Gibss sampling method, we provided an algorithm for estimating the parameters of the model and tested the model performance over numerical simulations.

The fact that SVM performed as random classifier implies that individual level infection statuses may not be estimated by a linear relationship with respect to available features such as the number of infected contacts in the past.  
Our supplementary tests with radial basis function (RBF) SVMs, that is a non-linear SVM, yielded the same results. 
Although outperforming the standard machine-learning model SVM, the GCHMM did not accurately predict the individuals' statuses when family tests were conducted sparsely.
Increasing the number of sampling iterations or using other optimization techniques such as the dual annealing \cite{xiang1997generalized} may improve the performance. 
The inclusion of disease symptoms as observations when available is another potential solution and subject to future work. 

When families were tested on a daily basis, the model well revealed the status of each individual. 
The number of daily tests was 23\% of the population size (15 over 64), implying a reduction by a factor of 4 in the number of testing kits. 
This indicates the potential power of the model in reducing screening costs by leveraging group tests.
The model may also be used for early warning signals due to its particularly high performance at the early stages of the spread. 
As time progresses, more infected individuals are to be detected, explaining the lower performance.  


A limitation of the model is to assume a low infection rate.
Moreover, the family or friendship network is assumed to be known at every time step. 
The estimation of the network via phone cellular data is a next step towards a realistic implementation of the model. 

\section{Appendix}
To illustrate the Gibbs sampling process, consider the graph shown in Figure \ref{myfigur}, its joint distribution is as follows:
\begin{equation}
\begin{aligned}
P(\bm{X,Y})=&P(X_{1,0})P(X_{2,0})P(X_{3,0})
P(X_{1,1}|X_{1,0},X_{3,0})P(X_{2,1}|X_{2,0})\cdot
\\&P(X_{3,1}|X_{1,0},X_{3,0})P(X_{1,2}|X_{1,1},X_{3,1})
P(X_{2,2}|X_{2,1},X_{3,1})\cdot
\\&P(X_{3,2}|X_{1,1},X_{2,1},X_{3,1})P(Y_{1,0}|X_{1,0})P(Y_{2,0}|X_{2,0})\cdot
\\&P(Y_{3,0}|X_{3,0})P(Y_{1,1}|X_{1,1})P(Y_{2,1}|X_{2,1})P(Y_{3,1}|X_{3,1})\cdot
\\&P(Y_{1,2}|X_{1,2})
\end{aligned}
\end{equation}
\tikzstyle{state}=[shape=circle,minimum size = 0.05cm, inner sep=.4pt,draw]
\tikzstyle{observation}=[shape=circle,minimum size = 0.05cm, inner sep=.4pt,draw,fill=gray!50]
\tikzstyle{edge}=[->,thick]

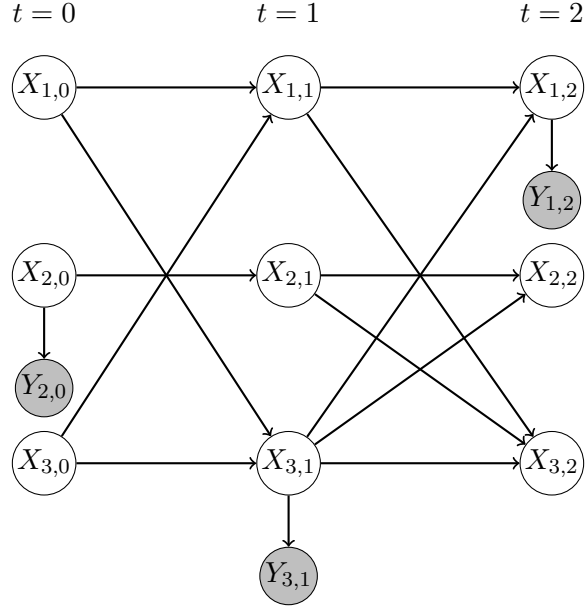
\begin{figure}[!htbp]
    \begin{center}
    \begin{tikzpicture}[]
    \node               at (-1.5,6.5) {$t=0$};
    \node[state] (s2) at (-1.5,3) {$X_{2,0}$};
    \node[state] (s3) at (-1.5,5.5) {$X_{1,0}$};
    
    \node[observation] (s4) at (-1.5,1.5) {$Y_{2,0}$};
    \node[state] (s5) at (-1.5,0.5) {$X_{3,0}$};
    
    \node               at (1.75,6.5) {$t=1$};
    \node[state] (s7) at (1.75,5.5) {$X_{1,1}$};
    \node[state] (s9) at (1.75,3) {$X_{2,1}$};
    \node[state] (s11) at (1.75,0.5) {$X_{3,1}$};
    \node[observation] (s12) at (1.75,-1) {$Y_{3,1}$};
    \node               at (5.25,6.5) {$t=2$};
    \node[state] (s13) at (5.25,5.5) {$X_{1,2}$};
    \node[observation] (s14) at (5.25,4) {$Y_{1,2}$};
    \node[state] (s15) at (5.25,3) {$X_{2,2}$};
    \node[state] (s17) at (5.25,0.5) {$X_{3,2}$};
    \draw (s3)edge[edge](s7);
    \draw (s3)edge[edge](s11);
    \draw (s5)edge[edge](s7);
    \draw (s7)edge[edge](s17);
    \draw (s11)edge[edge](s13);
    \draw (s9)edge[edge](s17);
    \draw (s11)edge[edge](s15);
    
    \draw (s7)edge[edge](s13);
    \draw(s13)edge[edge](s14);
    \draw(s2)edge[edge] (s4);
    \draw (s2)edge[edge](s9);
    
    \draw(s9)edge[edge] (s15);
    \draw(s5)edge[edge] (s11);
    \draw (s11)edge[edge](s12);
    \draw(s11)edge[edge](s17);
    \end{tikzpicture}
    \end{center}
    \label{myfigur}
    \caption{An example Graph coupled HMM}
\end{figure}$\newline$
To sample $X_{1,1}$ given current values of other variables, we use conditional probability definition:
\begin{equation}
    P(X_{1,1}\mid \{\bm{X,Y}\setminus X_{1,1}\})=\frac{P({\bm{X,Y}})}{\sum_{X_{1,1}=1}P(\bm{X,Y})}
\end{equation}
If we use (8) in (9), only the terms including $X_{1,1}$ will remain in the fraction. So we have:
\begin{equation}
\resizebox{\textwidth}{!}{%
  $\begin{aligned}
   & P(X_{1,1}|\{X,Y\}\setminus X_{1,1})=  {}\\
   & \frac{
       P(X_{1,1}|X_{1,0},X_{3,0})P(X_{1,2}|X_{1,1},X_{3,1})P(X_{3,2}|X_{1,1},X_{2,1},X_{3,1})P(Y_{1,1}|X_{1,1})
      }{
        \sum\limits_{X_{1,1}=0,1}^{}P(X_{1,1}|X_{1,0},X_{3,0})P(X_{1,2}|X_{1,1},X_{3,1})P(X_{3,2}|X_{1,1},X_{2,1},X_{3,1})P(Y_{1,1}|X_{1,1})
     }
  \end{aligned}$%
}
\end{equation}
By setting $X_{1,1}$ equal to $1$ in (10), the parameter $\lambda_1$ is obtained, where:

\begin{equation}
    \lambda_1=\frac{P(X_{1,1}=1|\{X,Y\}\setminus X_{1,1})}{(P(X_{1,1}=0|\{X,Y\}\setminus X_{1,1})+P(X_{1,1}=1|\{X,Y\}\setminus X_{1,1}))}
\end{equation}
Take the event $X_{1,1}=1$ as success, sampling a new value for $X_{1,1}$, is the same as sampling from a binomial distribution with the success rate $\lambda_1$. This means we generate a sample $d$ uniformly
from the interval [0, 1]. We then partition the interval into 2 subintervals: [0, $\lambda_1$), [$\lambda_1$, 1]. If $d$ is in the first interval, then the sampled value equals 0, otherwise it equals 1 [13]. Having sampled $X_{1,1}$, we now sample $X_{2,1}$, given this newly sampled value for $X_{1,1}$.

\printbibliography
\end{document}